# Human-AI Collaboration in Software Engineering: Lessons Learned from a Hands-On Workshop

Muhammad Hamza
Software Engineering
LUT University
Lahti, Finland
muhammad.hamza@lut.com

Dominik Siemon
Software Engineering
LUT University
Lahti, Finland
dominik.siemon@lut.fi

Muhammad Azeem Akbar
Software Engineering
LUT University
Lahti, Finland
azeem.akbar@lut.fi

Tahsinur Rahman
Software Engineering
LUT University
Lahti, Finland
tahsinur.rahman@lut.fi

## ABSTRACT

This paper investigates the dynamics of human-AI collaboration in software engineering, focusing on the use of ChatGPT. Through a thematic analysis of a hands-on workshop in which 22 professional software engineers collaborated for three hours with ChatGPT, we explore the transition of AI from a mere tool to a collaborative partner. The study identifies key themes such as the evolving nature of human-AI interaction, the capabilities of AI in software engineering tasks, and the challenges and limitations of integrating AI in this domain. The findings show that while AI, particularly ChatGPT, improves the efficiency of code generation and optimization, human oversight remains crucial, especially in areas requiring complex problem-solving and security considerations. This research contributes to the theoretical understanding of human-AI collaboration in software engineering and provides practical insights for effectively integrating AI tools into development processes. It highlights the need for clear role allocation, effective communication, and balanced AI-human collaboration to realize the full potential of AI in software engineering.

## 1 Introduction

The advent of artificial intelligence (AI) in software engineering has ushered in a new era of collaborative development, enabling new joint value creation between humans and AI systems [1]. AI has been defined as "a system's ability to correctly interpret external data, to learn from such data, and to use those learnings to achieve specific goals and tasks through flexible adaptation" [2]. The integration of AI in various domains has been the subject of extensive research and practical application [3]. Specifically in software engineering, AI tools have shown promise in automating mundane tasks, improving decision-making, and fostering innovative solutions [4] [5]. However, such systems can also have a negative impact on software engineering practices, thus affecting the whole process of how software is developed.

The launch of the "Chat Generative Pre-Trained Transformer" commonly known as ChatGPT, by OpenAI in late 2022 has garnered global interest in AI [6]. As of the end of January 2023, ChatGPT has reached a milestone of over 100 million monthly active users, achieving this feat within just two months of its launch. Figure 1 shows the rising trend of the term "ChatGPT" reflected in Google Searches. ChatGPT utilizes deep learning with extensive datasets for realistic conversation generation [7]. Enabled by generative AI, ChatGPT can create text, music, and images, and integrate data from various sources for analysis [8] that can be hard to differentiate from human-generated content.

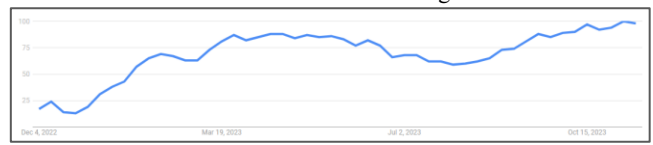

**Figure 1: Search trend of the term "ChatGPT" On Google.**

Generative AI has precipitated a spectrum of scientific implications and advancements across diverse sectors such as technology, business, education, healthcare as well as art and humanities [9]. Similarly, collaboration between humans and AI is crucial for leveraging the advantages of generative AI technologies such as ChatGPT [10]. For instance, in educational settings, instructors can employ ChatGPT in scientific teaching, ensuring a critical evaluation of AI-generated materials before their integration into curricula [11]. In the realm of problem-solving, generative AI can assist in ideation and the development or enhancement of solutions. In healthcare, ChatGPT has shown promise in supporting medical practices and offering insights in public health scenarios [12].

However, the collaboration of human AI in the context of software engineering is less explored. The role of AI in software development has evolved from being a mere tool to a collaborator, capable of adapting to specific roles such as senior backend developer, python expert, or software tester. Therefore, this research study explores the dynamics of human-AI collaboration in software engineering by drawing insights from a hands-on workshop where participants actively engaged with ChatGPT. The workshop provided a unique platform to observe and analyze how software engineers of varying expertise levels interact with AI in real-time development scenarios.

The rest of the paper is organized as section 2 discusses the related work. Similarly, in section 3, we delve into the methodology for conducting the workshop. Section 4 discusses the results of the workshop and section 5 discusses the threat to validity. Finally, section 6 concludes the study results.

## 2  Related Work

Since the advent of AI particularly large language models such as ChatGPT, various research studies have been conducted to address the different aspects of generative AI in different domains.

Marta Montenegro-Rueda et al [13]. highlight the positive influence of ChatGPT on educational processes, emphasizing the need for teacher training to maximize its effectiveness. Another study conducted by Jaber et al. [14], highlighted the diverse applications of ChatGPT in software development, including programming assistance, bug fixing, and enhancing the 3D printing process, underscoring its potential as a transformative tool in this field while noting the need for addressing its limitations and ethical concerns. Similarly, Sajed Jalil et al. [15] evaluate the performance of ChatGPT in a software testing curriculum, revealing that while it correctly or partially answers 55.6% of questions, its self-assessed confidence is often misaligned with the accuracy of responses, highlighting both the potential and limitations of using ChatGPT in educational settings. Another study conducted by Islam et al. [16] discusses the potential benefits and drawbacks of ChatGPT in academic contexts, emphasizing its capabilities in aiding research, education, and skill development while also cautioning about challenges related to bias, academic integrity, and the reduction of personalization and critical thinking skills.

A study conducted by Quanjun Zhang et al. [17] presents an in-depth analysis of the capabilities of ChatGPT in automated program repair. It demonstrates the ability of ChatGPT to outperform existing models by fixing a significant number of bugs, highlighting its potential for software engineering tasks while also noting the importance of prompt engineering and the challenges in evaluating black-box models like ChatGPT. Similarly, Ahmad et al. [18] explore the integration of ChatGPT in software architecture processes. It emphasizes the potential of human-bot collaboration in enhancing architectural analysis, synthesis, and evaluation, while also discussing the socio-technical challenges and the need for human oversight to ensure the consistency and ethical compliance of the generated architectural artifacts. A study conducted by Naomi S. Baron [19] discusses the integration of ChatGPT into software development, highlighting its benefits for coding efficiency and innovation, while also cautioning about the need for careful, informed approaches due to challenges in ensuring code quality and ethical considerations. Surameery & Shakor. [20] explore the potential of ChatGPT in debugging software. They highlight the strengths of ChatGPT in debugging assistance, bug prediction, and explanation while emphasizing the importance of integrating it with traditional debugging tools for more effective bug identification and resolution in software development. Nacthalia et al. [21] conducted an empirical comparison between ChatGPT and both novice and expert programmers. The study, focusing on solving LeetCode contest problems, found that ChatGPT performs better than novice programmers in easy and medium problems, but does not outperform expert programmers, highlighting the nuanced relationship between AI and human performance in software engineering tasks. Similarly, Bera et al. [22] examine the potential role of ChatGPT as a virtual team member in agile development teams. They explored how ChatGPT could provide generic knowledge on agile methods, give tailored advice, and even execute development work based on project data. The paper also analyzes Twitter sentiments related to ChatGPT's use in agile development and conducts an empirical experiment to assess the performance of ChatGPT in tasks typically done by an agile coach or Scrum master.

However, to the best of our knowledge, no study has been conducted to evaluate the role of generative AI as a human-AI collaborator. To this end, this study reports the lessons learned by conducting a workshop based on human-AI collaboration.

## 3  Research Method

This study was conducted within the broader context of examining the collaboration between humans, specifically software engineers, and artificial intelligence in software development. We employed a workshop-based research approach, aligning with the guidelines outlined by Thoring et al. [23]. Workshops are recognized as a valuable method in design science, action research, case-study-based research and action design research contexts [23]. Our focus was primarily on the evaluation of behavioral aspects of professional software engineers, with an emphasis on the interaction dynamics and collaborative processes between the participants and AI technologies. The workshop was designed to gather qualitative data relevant to our research objectives, which are how software engineers work together with ChatGPT to develop software. The workshop was prepared with components such as workshop facilitation tools, briefing of the moderators, and the creation of workshop materials (e.g., presentation slides) to ensure an understanding of the opportunities and characteristics that arise with working with ChatGPT. During the workshop, activities and discussions were structured to elicit insights into the collaborative practices and challenges faced by software engineers working with AI. The workshop took place in May 2023 in Finland as part of a two-day event on generative AI in software engineering, with participants from the industry with varying experiences organized by researchers and practitioners. In total, 22 software engineers took part in the workshop, with an average of 10 years of experience in software engineering. 3 participants were female. The whole procedure is depicted in the figure 2.

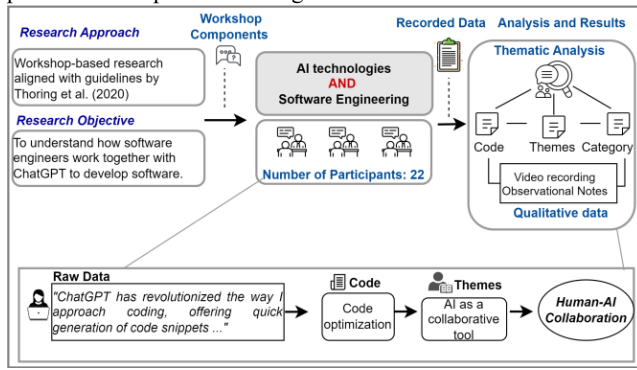

**Figure 2: Research Method**

Following the guidelines provided by Thoring et al. [23], our workshop had the following *(1) focus definition:* the workshop's focus was to evaluate the dynamics of human-AI collaboration in software engineering. The workshop aimed to observe and analyze interactions between software engineers of varying expertise levels and ChatGPT in real-time development scenarios. *(2) Role allocation:* The researcher who led the workshop gave instructions to the participants who were the subjects of the study. We separated the roles of facilitation and participant to maintain objectivity and integrity in the research process. *(3) Triangulation:* To ensure a comprehensive understanding of the collaborative dynamics, we employed a triangulation approach in our research. We combined various research methods and compared data from multiple sources. This allowed us to cross-validate findings and enhance the reliability of our observations. *(4) Transparency:* We followed established qualitative data analysis methodologies to ensure transparency. Furthermore, the data will be published in data repositories. *(5) Reflection:* Post-workshop, we critically reflected on the evaluation process, aiming to identify main insights regarding the utility and effectiveness of our methodology. This also involves, how specific data collection procedures benefited the understanding and knowledge creation [23].

The workshop began with a 25-minute instructional presentation on utilizing ChatGPT in software engineering, followed by a general introduction (e.g., presenting current research) to collaborating with AI, highlighting strategies for effective human-AI collaboration. This includes engaging in interactive discussions, facilitating turn-taking, assigning roles and responsibilities, encouraging the challenging of ideas and requesting specific justifications, as well as building a shared understanding of tasks and processes. The workshop then transitioned into a hands-on lab session, where participants applied these concepts in real-time, working with ChatGPT for coding over a period of two and a half hours. This practical segment was designed to mirror real-world software development scenarios, emphasizing interactive engagement with AI. The workshop concludes with a feedback and reflection phase, allowing participants to share experiences and insights, focusing on the nuances of AI collaboration in a professional context.

The entire workshop session was video recorded for comprehensive data collection. Additionally, one of the authors, who was leading the workshop, took notes, capturing behavioral observations, responses to open-ended questions, and other relevant interactions.

Finally, the recorded video was transcribed. We followed the thematic analysis approach to analyze the recording of the workshop, employing a multi-tiered coding process: deriving codes from raw data, mapping them into sub-themes, and subsequently themes for synthesizing these categories into overarching themes [24]. The whole procedure is depicted in the figure 2.

## 4 Results and Discussion

In this section, we describe the results obtained from the rigorous thematic analysis. We derived codes from the raw data and mapped them into sub-themes and themes. We derived four main themes i.e., human-AI collaboration, AI capabilities for SE, AI as a tool limitation, and adoption and learning process. The results are described as subsequent.

### 4.1 T1: Human- AI collaboration

The Human-AI Collaboration theme reflects an evolving perspective on artificial intelligence in the realm of software development. It emphasizes the transition of AI from a mere tool to a collaborative partner. This theme suggests a more integrated and interactive approach to using AI like ChatGPT in software engineering tasks. Participants learn to engage with the AI not just to execute predefined tasks, but also to contribute to the creative and problem-solving aspects of programming projects. This approach reflects a significant shift in how human professionals perceive and utilize AI capabilities in enhancing their software engineering workflows. As stated by one of the participants: *"Collaborating with AI has this group phenomena and group effects that you have in human-human collaboration also can be transferred to human-AI collaboration and this is actually also something I want to find out today as well".*

This theme emerged by different codes i.e., AI as a collaborative tool, defining and reminding the roles for AI, Iterative communication, and interactive learning and knowledge sharing.

**C1: AI as a collaborative tool:** The workshop demonstrated that AI, specifically tools like ChatGPT, can act as more than just tools; they can be collaborative partners. Participants engaged with AI to solve problems, optimize code, and even brainstorm ideas, reflecting a shift from viewing AI as a mere assistant to a collaborator. As stated by one of the participants: *"GPT did optimize the code. We will use ChatGPT as a collaboration partner in software engineering. Was it a collaboration? Yes, yes, definitely a collaboration, and with the other development too, we have a collaboration"*. Thus, participants from the workshop stated that AI such as ChatGPT can be used as a collaborative tool in the several domains of software engineering and showed high engagement with ChatGPT.

**C2: Defining and reminding the roles:** this code emphasizes assigning specific, complementary roles to both AI systems and human participants. Participants did tell ChatGPT to take up a specific role (Senior Backend Software Developer, Export in databases, Python expert, Front end developer, software tester and debugger, etc.). This approach not only enhances efficiency by delineating clear responsibilities but also optimizes the collaborative process by leveraging the unique strengths of each participant. AI, proficient in tasks like data analysis and executing repetitive tasks, complements human skills in creative thinking and complex problem-solving. Such clear role demarcation fosters a more integrated and effective team dynamic, allowing for a seamless fusion of human ingenuity and AI's computational power, leading to innovative solutions in fields like software development. As stated by one of the participants: *"In human-AI collaboration, you should define clear roles and for me, it was always OK really very basic things go to VS Code, do this and do this [...] it really helps you to know your collaborator which is ChatGPT better"*.

**C3: Effective interaction with AI:** effective and iterative communication with ChatGPT is essential for ensuring accurate and relevant responses, as it operates on the principle of "garbage in, garbage out". Clear and precise input reduces ambiguity, crucial in high-stakes fields of software engineering and medicine. This clarity also aids in the AI's learning process, enhancing its response quality over time. Moreover, it ensures a better user experience, making the AI more accessible and inclusive to diverse users. This iterative communication is key in collaborative AI-human tasks. The quality of GPT's output heavily depends on the clarity and specificity of the input. As stated by one of the participants: *"So if you want to do something really like really complex, then do something complex [...] Choose the programming language that you are familiar with and have effective interaction with AI to get most out of it"*.

**C4: Interactive learning and knowledge sharing**: this focuses on the importance of documenting interactions with AI for learning purposes and the value of sharing experiences and strategies among participants. This practice helped participants reflect on their experiences, understand what worked or did not, and provide a valuable resource for future reference. As stated by one participant *"I want you to document the process... So, document your process as good as possible"*.

## 4.2  T2: AI Capabilities for SE

In the realm of software engineering, ChatGPT offers a broad spectrum of capabilities identified during the workshop. The participants of the workshop highlighted and demonstrated different capabilities of ChatGPT that are grouped into this theme.

**C5: Code Generation and Assistance:** our participants confirmed that ChatGPT can generate code snippets, assist in writing functions, and offer syntax suggestions for various programming languages based on the specifications, helping them in coding tasks. According to one of the participants *"ChatGPT has revolutionized the way I approach coding, offering quick generation of code snippets that save time and enhance productivity"*. However, most experienced participants stated that ChatGPT can act as novice programmers but for complex tasks, human cognitive abilities are still required as stated by participants *"It cannot solve complex problems that necessitate human intuition, creativity, and deep contextual understanding [....]"*.

**C6: Documentation and Explanation:** ChatGPT can assist in drafting technical documents, user manuals, and code comments, enhancing clarity and accessibility. Additionally, ChatGPT simplifies technical jargon and provides analogies and examples to explain intricate programming concepts, making it easier for both experts and novices to understand. Most of the participants stated that it assists in drafting and executing the business requirements as stated *"Creating documentation is more efficient with ChatGPT. It helps in breaking down complex programming concepts into easy-to-understand parts"*.

**C7: Programming language versatility**: ChatGPT can assist in developing software in different programming languages making it more robust as a software engineering tool. Our participants were assigned the same task to do with different programming languages. Participants were impressed by ChatGPT's ability to understand multiple programming languages, highlighting its versatility as a coding assistant as stated by one of the participants *"By deeply engaging with ChatGPT, I appreciate how ChatGPT's understanding of different programming languages works as an assistant"*.

**C8: Code review and optimization:** ChatGPT's undeniable capabilities can review code developed by the novice or experienced practitioner. The practitioners were required to upload their projects and get their review by ChatGPT. This helps them review and optimize their existing codes. However, most of the participants were more concerned about privacy. As stated by one of the participants *"It helped me review my code and I was able to optimize it. However, it still has some limitations to understanding it"*.

**C9: Testing and quality assurance:** ChatGPT aids in testing (e.g., unit testing, integration testing) and quality assurance by automating test case generation and script writing, enhancing test efficiency. Additionally, ChatGPT offers suggestions for code quality improvement and adherence to coding standards. Participants of the workshop stated that testing small applications with ChatGPT can be helpful. However, for larger projects one still needs to understand its limitations.

### 4.3 T3: Technical Challenges and Limitations of AI

While GPT models offer significant benefits in software engineering, there are notable limitations to consider. Our participants in the workshop reported several concerns regarding the use of ChatGPT. We listed those codes in this theme.

**C10: Code Security:** One of the critical limitations of GPT models in software engineering is their potential inconsistency in adhering to the latest security best practices and coding standards. While GPT can generate code based on patterns it has learned, it doesn't inherently understand the security implications of these patterns. GPT-based tools can generate code efficiently, but this code may inadvertently contain security vulnerabilities such as susceptibility to SQL injection, cross-site scripting, or buffer overflows. Additionally, the code produced might not comply with industry-specific standards, leading to compliance issues in sectors that have stringent coding guidelines. As stated by one of the practitioners *"There's also a risk of overreliance on AI for security decisions, which could lead to a lapse in essential security due diligence normally conducted by human experts"*. To mitigate these risks, it's crucial to have human oversight, ensuring that experts review and validate GPT-generated code, especially for critical applications. However, code security for an application requires careful management and the continued involvement of human expertise to ensure compliance and safety.

**C11: Inability to debug:** the inability of ChatGPT to test or debug code is a significant limitation in their application to software engineering. While it can suggest potential fixes and improvements, but cannot execute or validate these suggestions, necessitating continuous human intervention for debugging. This limitation becomes particularly evident in complex systems, where understanding the interactions of code components and their behavior under various conditions is crucial. As stated by one of the participants *"Trust, but verify: GPT can propose, but only a developer's eye can truly diagnose and resolve"*.

**C12: Handling complex logic:** GPT models exhibit a significant limitation in processing and generating complex logical constructs and sophisticated algorithmic solutions. while GPT can proficiently handle routine coding tasks, its capacity to effectively tackle complex, nuanced programming scenarios is constrained, necessitating the intervention of experienced software engineers who possess the requisite expertise to navigate and resolve these high-complexity challenges. As stated by one of the practitioners *"In the realm of complex algorithms and intricate logic, GPT serves as a guidepost, yet the discernment and expertise of the seasoned software engineer remain indispensable"*.

### 4.4 T4: Adoption and Learning Process

Using ChatGPT involves a progressive learning process, starting with basic interactions and understanding the AI's capabilities and limitations. As users gain experience, they learn to craft more effective prompts and understand the importance of ethical and responsible use.

**C13: Learning curve:** The participants of the workshop initially spent time understanding ChatGPT's capabilities. This learning curve is crucial for effective utilization, indicating the importance of familiarizing oneself with AI tools to enhance collaboration. Less experienced participants went through an onboarding process, discussing capabilities and tasks with ChatGPT. This step was essential for them to effectively integrate ChatGPT into their workflow.

**C14: Social Loafing:** Social loafing on ChatGPT occurs when users depend excessively on AI for answers, reducing their intellectual engagement. This reliance can lead to diminished critical thinking and problem-solving skills, as users may not actively process information or develop their ideas. ChatGPT, while efficient and informative, should ideally complement rather than replace human effort. Professionals initially relied heavily on ChatGPT, putting in less effort themselves. This reliance demonstrates both the perceived efficiency of AI assistance and the potential risk of reduced human effort in AI-collaborative environments as observed in the workshop.

All the results and their mapping are presented in figure 3.

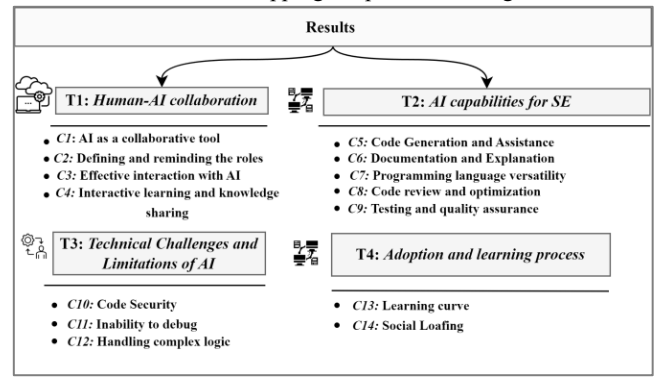

**Figure 3: Thematic analysis results.**

## 5 Implications

Based on the results of the workshop's thematic analysis, the implications for both theory and practice in the areas of software engineering and human-AI collaboration are diverse and significant.

In terms of theoretical implications, the emergence of the human-AI collaboration theme marks a paradigm shift in how AI is perceived and used in software engineering Ebert and Louridas [4] Traditionally viewed as a tool (for example such as GitHub Copilot), the evolving role of AI as a collaborative partner suggests a more integrated approach, where AI such as ChatGPT contributes not only to routine tasks, but also to creative and problem-solving aspects of projects (Seeber et al. [25]. The participants reported that they relied partially on ChatGPT's "creativity" to, for example, create, a unique interface or implement a function without giving specific requirements, i.e., leaving the solution to ChatGPT. Furthermore, the dynamics within the collaboration, like turn-taking or challenging ChatGPT to justify specific decisions, change the nature of the work of software engineers. This shift requires a re-evaluation of existing theories of team dynamics and collaboration in software engineering, considering the unique capabilities, roles, and interactions presented by AI partners. Insights from the workshop, such as the need for clear role delineation (Siemon et al. [26] and effective communication with AI, point to new theoretical frameworks that integrate these elements into the human-AI collaboration model. In addition, the way turn-taking was achieved as well as creating a shared understanding of the task contributes to current research in the field (Hevner and Story [27]; Siemon et al. [28]

In practice, the workshop's findings have several direct implications. Recognizing AI as a collaborative tool, rather than just an assistant, changes the way software engineers approach their work [4]. By assigning specific roles to AI systems, such as a Python expert or a software tester, teams can create an efficient hybrid intelligence, with AI and humans focusing on designated areas of the software engineering process, thus creating a cohesive team. This approach optimizes efficiency and encourages innovative solutions in software development. In addition, the need for iterative and effective communication with AI, as highlighted by the results, emphasizes the importance of clear and precise input in software engineering contexts. This clarity not only improves the quality of the AI's response but also improves the overall user experience, making AI tools more accessible and inclusive.

In addition, the workshop highlighted AI capabilities specific to software engineering, such as code generation, assistance, and versatility in multiple programming languages which further contributes to the theoretical understanding of how AI can be leveraged in software engineering. These capabilities, when combined with human oversight, particularly in areas such as code security and debugging complex logic, can significantly improve the efficiency and quality of software development. However, it also underlines the importance of maintaining human expertise, especially in tasks involving complex logic and security considerations.

The findings also point to the need for a careful adoption and learning process when integrating AI into software engineering practices in line with theories such as human-in-the-loop Gronsund and Aanestad [29] or hybrid intelligence. The learning curve associated with AI tools such as ChatGPT is an important consideration for practitioners, highlighting the importance of understanding the capabilities and limitations of these tools. In addition, the phenomenon of social loafing, where over-reliance on AI can lead to reduced human effort and critical thinking, highlights the need for balanced AI-human collaboration as shown in other research related to human-AI collaboration Siemon and Wank [30].

## 6    Threat to validity

Various potential threats could affect the validity of the study findings. The relevant threats are broadly categorized across internal, construct, and external validity [31] .

**Internal validity:** Internal validity is the extent to which particular factors affect the methodological rigor. In this study, the first threat to the internal validity is the data collected during the workshop session. This threat has been mitigated by conducting pilot sessions to ensure the understandability and reliability of the expert's discussion.

**Construct validity:** Construct validity is the extent to which the study constructs are well-defined and interpreted. In this study, the workshop participants' perceptions of human-AI collaboration adaptability and relevant challenges are the core constructs. The verifiability of constructs is a known limitation of this types of study. It could be inferred from the research method efficacy and from the evidence that the study findings are presented based on the data collected using the selected research method [32]. Therefore, we explicitly discussed the step-by-step process of the research method, the defined categories supported with quotes from the workshop participants, and our observations in section 3. It exhibits how the systematically defined research protocol and reported findings support the verifiability of the study constructs.

**External validity:** External validity refers to broadly generalizing the study findings in other contexts. In this study, the sample size and sampling approach may not provide a strong foundation to generalize the findings. However, it is known that human-AI collaboration, particularly using ChatGPT, are new research areas and not in practice at a high level. We tried to mitigate this threat by using all possible sources to approach the potential population and arranging a constructive workshop. Moreover, we plan to extend this study in the future with a large data sample from multiple data sources (i.e. mining the Q&A platforms, conducting industrial surveys and interviews).

## 7    Conclusion and Future Work

This study's exploration of human-AI collaboration in software engineering, specifically using ChatGPT, has highlighted the changing dynamics and collaboration in the field. The findings reveal a significant evolution in the role of AI - from a mere tool to an active collaborator. This transition suggests a more nuanced approach to AI integration in software engineering, where the role of AI extends to creative problem-solving and interactive learning. The study highlights the importance of a clear role definition, effective communication, and a balanced approach to human-AI collaboration, highlighting AI's capabilities in tasks such as code generation, documentation, and understanding multiple programming languages. Future research should focus on further refining human-AI collaboration practices in software engineering. This includes investigating the long-term impact of collaborating with AI in all software engineering activities, software project management, overall software quality, and related software engineering activities. In addition, it is crucial to address the challenges identified, such as code security, and the limitations of AI in debugging and handling complex logic. Research could also explore how to effectively manage the learning curve associated with AI tools and mitigate the risk of social loafing, as the use of AI will become more common practice and has just started to be more intensively used. Finally, extending this research to other domains where human-AI collaboration is emerging will provide valuable cross-disciplinary insights and contribute to a broader understanding of the role of AI in different professional contexts.